\documentstyle[epsfig,12pt]{article}

\newcommand{\bce}{\begin{center}} 
\newcommand{\ece}{\end{center}}
\newcommand{\beq}{\begin{equation}}
\newcommand{\eeq}{\end{equation}}
\newcommand{\bea}{\vspace{0.25cm}\begin{eqnarray}}
\newcommand{\eea}{\end{eqnarray}}

\newcommand{\bk}{{\bf k}}

\newcommand{\bm}[1]{\mbox{\boldmath $#1$}}

\newcommand{\ba}{\begin{array}}
\newcommand{\ea}{\end{array}}

\newcommand{\ket}[1]{| {#1} \rangle}

\newcommand{\doublespace}{
    \renewcommand{\baselinestretch}{1.6}\large\normalsize}

\newcommand{\bkappa}{\mbox{\boldmath ${\kappa}$}}

\def\lsim{\mathrel{\rlap{\lower4pt\hbox{\hskip1pt$\sim$}}
    \raise1pt\hbox{$<$}}}         %less than or approx. symbol
\def\gsim{\mathrel{\rlap{\lower4pt\hbox{\hskip1pt$\sim$}}
    \raise1pt\hbox{$>$}}}         %greater than or approx. symbol

\def\beq{\begin{equation}}
\def\endeq{\end{equation}}
\def\arr{\begin{eqnarray}}
\def\endarr{\end{eqnarray}}

\makeindex

  \advance\oddsidemargin  by -1in
  \advance\evensidemargin by -1in
  \advance\topmargin      by -0.5in
\begin{document}

\vspace{2.0cm}

\begin{flushright}
% ITEP(Ph)-2008-08
\end{flushright}

\vspace{1.0cm}

\begin{center}
{\Large \bf 
Current non-conservation effects in ultra-high energy neutrino 
interactions} 

\vspace{1.0cm}

{\large\bf R.~Fiore$^{1 \dagger}$ and V.R.~Zoller$^{2 \ddagger}$}

\vspace{1.0cm}

$^1${\it Dipartimento di Fisica,
Universit\`a     della Calabria\\
and\\
 Istituto Nazionale
di Fisica Nucleare, Gruppo collegato di Cosenza,\\
I-87036 Rende, Cosenza, Italy}\\
$^2${\it
ITEP, Moscow 117218, Russia\\}
\vspace{1.0cm}
{ \bf Abstract }\\
\end{center}
The overall hardness scale of the 
ultra-high energy neutrino-nucleon  interactions is usually estimated as 
$Q^2\sim m_W^2$.
The effect of non-conservation of weak currents   pushes this scale up to 
the top quark mass squared and changes dynamics of the scattering  process.
   The Double Leading Log 
Approximation  provides simple and numerically accurate formula for  the 
top-bottom contribution to the total cross section  $\sigma^{\nu N}$.
Corresponding   correction to 
$\sigma^{\nu N}$ appears to be numerically large. It is  comparable 
with the leading contribution  evaluated in the 
massless quark approximation.

\doublespace

\vskip 0.5cm \vfill $\begin{array}{ll}
^{\dagger}\mbox{{\it email address:}} & \mbox{fiore@cs.infn.it} \\
^{\ddagger}\mbox{{\it email address:}} & \mbox{zoller@itep.ru} \\
\end{array}$

\pagebreak

%---------------------------------------------------

%-------------------------------------------

New ideas \cite{Becker} about the origin of neutrino  fluxes  from active galactic 
nuclei, gamma ray bursts or from decay of exotic heavy particles inspired
many publications  on the ultra-high energy (UHE) neutrino-nucleon total cross 
sections $\sigma^{\nu N}$ \cite{List1,List2,HJM,KK2003}. 
The UHE interactions correspond to neutrino  energy  
above $E_{\nu}\sim 10^8$ GeV, where the gauge boson exchange   probes the gluon 
density in the target  nucleon at 
very small values of  Bjorken $x$. The gluon density at small $x$ is known to be  
a rapidly rising function of 
$Q^2$. Its rise is  tamed, however,  by the propagator 
of the gauge boson which sets the  restriction \cite{List1,List2,HJM,KK2003}
\beq
Q^2\lsim m_W^2.
\label{eq:Q2W2}
\eeq
This value of $Q^2$ represents the overall hardness scale of the process 
induced by the light quark current, $m_q^2\ll Q^2$. 
The top-bottom current needs special care.  In this communication 
we show that the 
charged  current non-conservation (CCNC)  effect 
 pushes the hardness scale up to the top quark mass squared, $m_t^2$, 
and crucially changes dynamics of  the process 
\footnote{Preliminary results have 
been reported at the Diffraction 2010 Workshop \cite{UHE10}}. 

The differential  cross section for the neutrino-nucleon interactions is
 expressible 
 in terms of the
longitudinal, $F_L$, transverse, $F_T$, and left-right antisymmetric, $F_3$, 
structure functions. In standard notations it reads 
\beq
x{d\sigma^{\nu N}\over dxdQ^2}={G_F^2\over 2\pi}
\left({m_W^2\over m_W^2+Q^2} \right)^2
\left[(1-y)F_L+(1-y+{y^2\over 2})F_T +y(1-{y\over 2})xF_3\right],
\label{eq:DSDXDY}
\eeq
It is the
longitudinal structure function $F_L$
which is the carrier of   
 the CCNC effect. 
Indeed,
for longitudinal/scalar  W-boson with polarization vector 
$\varepsilon^L_{\mu}$ the vector or axial-vector  transition vertex 
$W\to t\bar b$ is 
$\propto 
\varepsilon^L_{\mu}J_{\mu}\propto\partial_{\mu}J_{\mu}\propto m_t\pm m_b.
$
Therefore,
$F_L$ which is 
$\propto \varepsilon^L_{\mu}T_{\mu\nu}\varepsilon^L_{\nu}$ 
provides a measure of the CCNC effect, here 
$T_{\mu\nu}$ stands for the imaginary part of the  forward  scattering Compton amplitude. 

\begin{figure}[h]
\psfig{figure=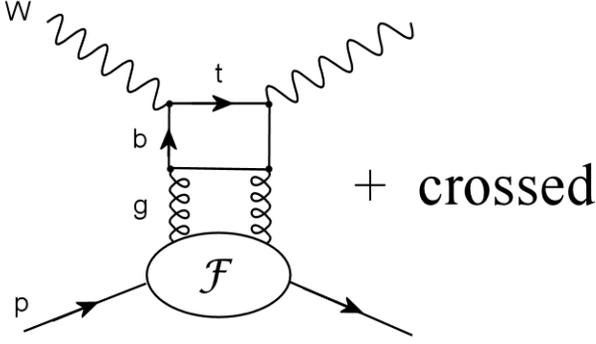,scale=1.0}
\vspace{-0.5cm}
\caption{Diagrammatic representation of the $\bkappa$-factorization
 formula (\ref{eq:1.1})} 
\label{fig:fig1}
\end{figure}

To the Leading Log$({1/x})$ Approximation, the operational 
definition of the differential glue for any 
target is provided by the $\bkappa$-factorization 
representation{\footnote{The differential cross section 
$d\sigma_{L,T}/dz d^{2}\bk$  to the lowest  order in  pQCD has 
been derived 
 in \cite{BGNPZ} }} 
corresponding to the gauge 
invariant sum of diagrams like that shown in Fig.~\ref{fig:fig1}
\bea
{d\sigma_{L}(x,Q^{2})\over dz d^{2}\bk }= 
{\alpha_{W}\over \pi}
 \int { d^{2}\bkappa 
\over \bkappa ^{4}}\alpha_{S}(q^{2})
{\cal F}(x,\bkappa ^{2})\left(V_S+A_S+V_P+A_P\right)
\label{eq:1.1}
\eea
where $\alpha_W=g^2/4\pi$, $g^2=G_Fm_W^2/\sqrt{2}$ and  
${\cal F}$ stands for  the differential gluon density 
$${\cal F}(x,\bkappa ^{2})=
{\partial G(x,{\bm \kappa}^2)\over \partial \log\kappa^2}.$$ 
We denoted by $ \bkappa$ and $\bk$    the gluon and $t$-quark transverse  
momenta, respectively, and by  $z$    
 the  fraction of the light--cone momentum
of the $W$  carried by the top quark. 

At small $x$
it is legitimate to discuss the $\nu N$-scattering in the laboratory frame
in terms of interactions with the target of the $q\bar q\prime$-pair which the 
light-cone W-boson transforms into at large upstream distances.
  The axial-vector $A_S$ and  vector term 
 $V_S$ 
describe the interaction with the target  of the  
quark-antiquark $\ket{t\bar b}$ state
 with 
the angular momentum $L=0$ (S-wave), 
\bea
V_S(m_t,m_b)=
{g_V^2\over Q^2}\left\{2Q^2z(1-z)
+(m_t-m_b)\left[(1-z)m_t-zm_b\right]\right\}^2\nonumber\\
\times\left({1   \over  \bk^{2}+\varepsilon^{2}}-
{1 \over  (\bk-\bkappa )^{2}+\varepsilon^{2}}\right)^{2},\nonumber\\
A_S(m_t,m_b)={g^2_A\over g_V^2} V_S(m_t,-m_b),
\eea
where
\beq
\varepsilon^2=z(1-z)Q^{2}+(1-z)m_{t}^{2}+zm_b^2.
\label{eq:1.3}
\eeq
In the charged current neutrino interactions $g_{A}=-g_{V}=-1$
and $m_b$ and $m_t$ stand for the bottom and the top
quark masses.
The two terms $A_P$ and $V_P$ 
 correspond  to the 
quark-antiquark states with the angular momentum $L=1$ (P-wave) and are given by
\bea
V_P(m_t,m_b)={g_V^2\over Q^2}(m_t-m_b)^2\left({\bk \over  \bk^{2}+\varepsilon^{2}} - 
{\bk-\bkappa  \over  (\bk-\bkappa )^{2}+\varepsilon^{2}}\right)^{2},\nonumber\\
A_P(m_t,m_b)={g^2_A\over g^2_V} V_P(m_t,-m_b).
\label{eq:1.33}
\eea
The P-wave component of the light-cone Fock state expansion for the 
longitudinal/scalar W-boson  arises
entirely due  to the current non-conservation. In the P-wave quark-antiquark 
state  
$\ket{q\bar q^{\prime}}$ either quark or antiquark has 
wrong helicity, the quark is right-handed or antiquark is left-handed. 
Normally, this configuration is suppressed as $m_q^2/Q^2$ but in our 
specific case $Q^2$ is
 limited (see Eq.(\ref{eq:Q2W2})) and  $m^2_q=m^2_t\gg Q^2$. Therefore, 
to describe correctly  the suppression of  wrong helicity states one needs 
more accurate treatment.
 We addressed this issue in \cite{FZCS1,FZCS2} and 
 quantified the CCNC effect in terms of the light cone wave functions
in the color dipole basis.
Here we reproduce and extend the results of \cite{FZCS1,FZCS2} making use 
of the momentum 
representation.

Consider first the P-wave component of Eq.(\ref{eq:1.1}). 
Separate the $\bkappa^{2}$-integration in (\ref{eq:1.1}) into 
the soft, 
$$\bkappa^{2} \ll \overline{k^2}\equiv\varepsilon^{2}+{\bm k}^{2}$$
 and hard, 
$$\bkappa^{2} \gsim \overline{k^2}, $$
regions of the gluon momentum. For soft gluons  upon the azimuthal integration
we get 
\bea
\int d\varphi\left({\bk \over  \bk^{2}+\varepsilon^{2}} - 
{\bk-\bkappa  \over  (\bk-\bkappa )^{2}+\varepsilon^{2}}\right)^{2}\simeq
2\pi\bkappa ^2{ \varepsilon^{4} + (\bk^2)^2 \over (\bk^{2}+\varepsilon^{2})^4}.
\label{eq:10.6}
\eea
In Eq.(\ref{eq:1.1}) the QCD running coupling $\alpha_{S}(q^{2})$ enters 
the integrand at the largest
relevant virtuality,
$q^2={\rm max }\{\overline{k^2},\bkappa ^{2}\}.$
To the Double Leading Log Approximation (DLLA) \cite{DLLA} one can take
$
q^{2}=\overline{k^2}$
and for soft gluons we arrive at the fully differential
distribution of the t-quark in $z$ and $\bk$,
 \bea
{d\sigma^P_{L}(x,Q^{2})\over dz d^{2}\bk }= 
\alpha_{W}(g_A^2+g_V^2){m_t^2\over Q^2}
 \alpha_{S}(\overline{k^2})
G(x,\overline{k^2})
{\varepsilon^{4} + 
(\bk^2)^2 \over (\bk^{2}+\varepsilon^{2})^4}, 
\label{eq:1.7}
\eea
where $m^2_b$ is  neglected compared to  $m^2_t$.
Then, going from ${d\sigma_{L}/dz d^{2}\bk }$ to the longitudinal
 structure function,
$$
F_L(x,Q^2) = {Q^2 \over 4\pi^2 \alpha_{W}}\sigma_{L}(x,Q^{2})
$$
we find  the soft gluon contribution to the P-wave component of $F_L$
\bea
F^P_{L}(x,Q^{2})\simeq
{m_t^2\over 2\pi^2}\int_{0}^1 dz
 \int d^{2}\bk 
\alpha_{S}(\overline{k^2})
G(x,\overline{k^2}){\varepsilon^{4} + 
(\bk^2)^2 \over (\bk^{2}+\varepsilon^{2})^4},
\eea
where 
\beq
x={{Q^2+M^2}\over {W^2+Q^2}}  
\eeq
and in the soft gluon approximation
\beq
M^2={{m_t^2+{\bk}^2}\over {z}}+
{{m_b^2+{\bk}^2}\over {1-z}}.
\eeq
The full $z$ integration can be 
separated into two domains $z_m<z<1$ and  $0<z<z_m$,
 where 
\beq
z_m=max\left\{0,\left[1-(m_t^2-m_b^2)/Q^2\right]/2\right\}.
\eeq
 The leading contribution to 
$F^P_L$ comes from 
\beq 
z\sim 1-{m_b^2\over m_t^2+Q^2},
\label{eq:zlead}
\eeq 
when the  t-quark carries
almost 100 per cent of the longitudinal W's momentum \cite{FZCS1}, 
so that for $z_m<z<1$ we
 can make a substitution 
$d\varepsilon^2 =-(Q^2 +m_t^2)dz$.  Then
\bea
F^P_{L}(x,Q^{2})\simeq
{m_t^2\over {m_t^2+Q^2}}\int_{m_b^2}^{\varepsilon_m^2}
{d\varepsilon^{2}\over \varepsilon^{2}} 
 {\alpha_S(\varepsilon^{2}) \over 3\pi}G(x,\varepsilon^2).
\label{eq:1.8}
\eea
Here the hardness scale for  $Q^2\leq m_t^2-m_b^2$ is
\beq
\varepsilon_m^{2}=m_t^2
\eeq
and for higher  $Q^2>m_t^2-m_b^2$ it is
\beq
\varepsilon_m^{2}={1\over 4}\left(Q^2+m_t^2+m_b^2\right)\left[1+
({m_t^2-m_b^2})/Q^2\right];
\label{eq:1.9}
\eeq

The CCNC also affects the  S-wave component of the longitudinal structure function,
  $$ F_L=F_L^P+F_L^S. $$ 
One can see that both  the  S-wave and P-wave parts of $F_L$
correspond to   very different  $z-$distributions.
The P-wave component is dominated by $z\sim 1$ while the 
S-wave term 
 integrated over $\bk$ has 
approximately uniform $z$-distribution.
A narrow peak in the S-wave $z$-distribution at $z\to 1$ 
rises to $\sim (m_t^2+Q^2)/m_b^2$ but its  width is
$\delta z\sim m_b^2/(m_t^2+Q^2)$ and this singularity
 does not affect the DLLA  estimate 
\bea
F_L^S(x,Q^2)\simeq {2\alpha_S(\overline{\varepsilon^{2}})
\over 3\pi}
G(x,\overline{\varepsilon^{2}}),
\label{eq:1.11}
\eea
where  
$$\overline{\varepsilon^{2}}\simeq (Q^2+2m_t^2)/4$$ 

We neglected here the contribution of  hard gluons 
to the proton longitudinal 
 structure function. Therefore, the DLLA gives  the lower estimate for $F_L$.

Then the contribution to $\sigma^{\nu N}$,
\beq
\sigma^{\nu N}=\int_{Q_0^2}^s dQ^2
\int_{x_t}^1 {dx}\left({d\sigma\over dxdQ^2}\right),
\label{eq:SIGMA}
\eeq
 coming from the absorption of longitudinal W-bosons
(we call it the CCNC contribution) can easily be estimated,
in Eq.(\ref{eq:SIGMA})  $x_t=(m_t^2+Q^2)/s$, $y=(m_t^2+Q^2)/xs$ and $s=2m_NE_{\nu}$.   At
$E_{\nu}=10^{12}\,\, GeV$ and  for the input  gluon density
 $G(x,k^2)$ specified in 
 \cite{IN2003} this contribution appears to be equal to 
\beq
\sigma_{CCNC}^{\nu N}\simeq 0.45\times 10^{-31}\,cm^2.
\label{eq:IVNIK}
\eeq
While for  the gluon density from \cite{GRV98}  we arrive at
\beq
\sigma_{CCNC}^{\nu N}\simeq 0.56\times 10^{-31}\,cm^2
\label{eq:GRV}
\eeq
For comparison, the frequently used approximation of
massless quarks gives
  the total cross section $\sigma^{\nu N}$  
which for different  input
gluon densities  varies   in a rather  wide range \cite{HJM}. 
For example, at $E_{\nu}=10^{12}\,\, GeV$,
\beq
0.2\times 10^{-31}\, cm^2< \sigma^{\nu N}
< 1.5\times 10^{-31}\,cm^2.
\label{eq:RANGE}
\eeq
Thus, the CCNC correction to the massless $\sigma^{\nu N}$ is comparable with  $\sigma^{\nu N}$.

At small Bjorken $x$ the unitarity/saturation effects enter the game \cite{Kancheli73, NZ75}. 
Corresponding  correction to 
$\sigma^{\nu N}$ was estimated in \cite{KK2003}
as a  $50$ per cent effect.
 Particularly, it was found that the account of the  unitarity  
 turns the charged current cross section 
$$\sigma^{\nu N}\simeq 1.\times 10^{-31}cm^2,$$ 
 at $E_{\nu}=10^{12}$ GeV  into 
$$\sigma^{\nu N}\simeq 0.5\times 10^{-31}cm^2.$$
In \cite{KK2003} the massless quark  approximation was used.
The unitarity  effect  is known to  depend on the hardness 
scale of the process.  The first higher twist correction  is usually  estimated as \cite{GLR}
$$\sim {\alpha_S(Q^2)\over Q^2} {G(x,Q^2)\over \pi R^2}.$$ 
It was noted  above that the CCNC hardness scale  is much ``harder'' 
than  the  corresponding  scale  for light flavors:
$m_t^2\gg Q^2\sim m^2_W.$
Therefore,  the unitarity  correction  to the CCNC component of 
$\sigma^{\nu N}$ is expected to be  much smaller.

Summarizing, it is shown  that in the UHE neutrino interactions  the higher twist corrections
brought about by  the non-conservation  of the  top-bottom current 
dramatically change the longitudinal structure function, $F_L$, Eqs.(\ref{eq:1.8}). We started
 with the $\bkappa$-factorization formula  for the  differential cross section 
$d\sigma_{L}/dz d^{2}\bk$ and derived simple and numerically accurate
DLLA expression for $F_L$. It is worth emphasizing that the appearance of the factor 
$
{m_t^2/(m_t^2+Q^2)}
$
in (\ref{eq:1.8})  is not a property of the interaction of 
$t\bar b$-dipole  with the target but
the property of the light-cone density of  
$t\bar b$-states \cite{FZCS1}. Only relative smallness of  $Q^2$ restricted by 
Eq.(\ref{eq:Q2W2}) prevents the contribution of the CCNC  term to 
$\sigma^{\nu N}$ from vanishing.  The rapidly rising  gluon density factor 
provides its additional enhancement.
We neglected  here the contribution of  hard gluons to  $F_L$. 
Therefore, the DLLA gives  the lower estimate for 
the CCNC contribution to  $\sigma^{\nu N}$. This contribution appears to be 
numerically large and  comparable
with $\sigma^{\nu N}$ evaluated in the massless quark  approximation.
Curiously, the  CCNC effect in its competition 
with massless calculations gains  momentum also  from  the unitarity suppression  
which is much stronger for the  massless component of $\sigma^{\nu N}$.

{\bf Acknowledgments.} Thanks are due  to 
G. Ciapetta for the help in preparation of the manuscript.
V.R.~Z. thanks B.G.~Zakharov for discussions and 
 the Dipartimento di Fisica dell'Universit\`a
della Calabria and the Istituto Nazionale di Fisica
Nucleare - gruppo collegato di Cosenza for their warm
hospitality while a part of this work was done.
The work was supported in part by the Ministero Italiano
dell'Istruzione, dell'Universit\`a e della Ricerca and  by
 the RFBR grant 09-02-00732.


\begin{thebibliography}{299}
\bibitem{Becker} J.K. Becker, Phys. Reports, {\bf 458}, 173 (2008).

\bibitem{List1} G.M. Frichter, D.W. McKay and J.P. Ralston,
 Phys. Rev. Lett. {\bf 74}, 1508 (1995);
 Erratum-ibid. {\bf 77}, 4107 (1996);
R. Gandhi, C. Quigg, M.H. Reno and I. Sarcevic, Astropart. Phys.{ \bf 5}, 81 (1996); 
Phys. Rev. {\bf D 58}, 093009 (1998);
 G. Parente and E. Zas, Proc. of the 7th Intern. Symp. on Neutrino Telescopes, Venice,
Italy, Feb. 1996, p. 499 (astro-ph/9606091);
G.C. Hill, Astropart. Phys. {\bf 6}  215 (1997);  
M.V.T. Machado,  Phys.Rev. {\bf D70}, 053008 (2004);
R. Fiore, L.L. Jenkovszky, A.V. Kotikov, F. Paccanoni and A. Papa, Phys.Rev. {\bf D73}, 
053012 (2006); 

\bibitem{List2}
M. Gl\"uck, S. Kretzer and  E. Reya, Astropart.Phys. {\bf 11}, 327 (1999);
J. Kwiecinski, Alan D. Martin and  A.M. Stasto, Phys.Rev. {\bf D59}, 093002 (1999); 
Yu Seon Jeong and  Mary Hall Reno, Phys.Rev. {\bf D81}, 114012 (2010); 
M. Gl\"uck, P. Jimenez-Delgado and  E. Reya, Phys.Rev. {\bf D81}, 097501 (2010). 

\bibitem{HJM}
E.M. Henley and  J. Jalilian-Marian, Phys.Rev. {\bf D73}, 094004 (2006).

\bibitem{KK2003}
K. Kutak, J. Kwiecinski, Eur.Phys.J. {\bf C29}, 521 (2003). 

\bibitem{UHE10} R. Fiore and V.R. Zoller 
``UHE neutrinos: current non-conservation, mass scales, saturation''
talk at Diffraction 2010, Otranto (Lecce), Italy, September 10 - 15, 2010.

\bibitem{BGNPZ} V. Barone, M. Genovese, Nikolai N. Nikolaev, E. Predazzi and
 B. Zakharov, Phys.Lett. {\bf B328}, 143 (1994). 




\bibitem{FZCS1} R. Fiore and V.R. Zoller,
JETP Lett. {\bf 87}, 524 (2008); 
``Full of charm neutrino DIS'',
in {\emph '08 QCD and High Energy Interactions},
Proc. of 43rd Rencontres de Moriond on QCD and Hadronic Interactions, 
La Thuile, Italy, 2008, e-Print: arXiv:0805.2090.  

\bibitem{FZCS2} 
R. Fiore and  V.R. Zoller, Phys.Lett. {\bf B681}, 32(2009); 
``Current non-conservation effects in ${\nu}$DIS diffraction''
in AIP Conf.Proc. {\bf 1105}, 304 (2009). 

\bibitem{DLLA} 
V.N. Gribov and L.N. Lipatov, Sov. J. Nucl. Phys. {\bf 15}, 438 (1972);
L.N.Lipatov, Sov. J. Nucl. Phys. {\bf 20}, 181 (1974); Yu.L. Dkshitzer,
Sov. Phys. JETP {\bf 46}, 641 (1977); G. Altarelli and G. Parisi, Nucl. Phys.
{\bf B126}, 298 (1977); R.G. Roberts, The structure of the proton. 
(Cambridge Univ. Press, 1990)

\bibitem{IN2003} I.P. Ivanov and N.N. Nikolaev, Phys. Rev. {\bf D65}, 054004 
(2003).

\bibitem{GRV98} M. Gl\"uck, E. Reya and A.Vogt, Eur. Phys. J. {\bf C5}, 461 
(1998). 

\bibitem{Kancheli73} O.V. Kancheli, Sov. Phys. JETP Lett {\bf 18}, 274 (1973).
\bibitem{NZ75} N.N. Nikolaev and V.I. Zakharov, Phys. Lett. {\bf  B55}, 397 (1975); 
Sov. J. Nucl. Phys. {\bf 21}, 227 (1975)

\bibitem{GLR} L.V. Gribov, E.M. Levin, M.G. Ryskin, Phys. Rep. {\bf 100}, 1
 (1981); A.H. Mueller, J. Qiu, Nucl. Phys. {\bf B 268}, 427 (1986).

\end{thebibliography}
\end{document}